\documentclass[runningheads]{llncs}
\usepackage[T1]{fontenc}
\usepackage{float}
\usepackage{amsmath}
\usepackage{amssymb}
\usepackage{graphicx}  
\usepackage{hyperref}
\usepackage{adjustbox}
\usepackage{etoolbox}
\usepackage{siunitx}
\usepackage{booktabs}
\usepackage{array}
\usepackage{multirow} 

\usepackage{graphicx,verbatim}
\begin{document}
\title{Cross-Attention fusion of MRI and Jacobian maps for Alzheimer’s Disease Diagnosis}
%

\author{Shijia Zhang, MBI\inst{1} \and Xiyu Ding, MS\inst{1} \and Brian Caffo, PhD\inst{1} \and Junyu Chen, PhD\inst{1} \and Xinyi Zhang, PhD\inst{1} \and Hadi Kharrazi, PhD\inst{1} \and Zheyu, Wang PhD\inst{1}}

\authorrunning{S. Zhang et al.}
\institute{Johns Hopkins University \\
    \email{wangzy@jhu.edu}}

\maketitle              
\begin{abstract}
Early diagnosis of Alzheimer’s disease (AD) is critical for intervention before irreversible neurodegeneration occurs. Structural MRI (sMRI) is widely used for AD diagnosis, but conventional deep learning approaches primarily rely on intensity-based features, which require large datasets to capture subtle structural changes. Jacobian determinant maps (JSM) provide complementary information by encoding localized brain deformations, yet existing multimodal fusion strategies fail to fully integrate these features with sMRI. We propose a cross-attention fusion framework to model the intrinsic relationship between sMRI intensity and JSM-derived deformations for AD classification. Using the Alzheimer’s Disease Neuroimaging Initiative (ADNI) dataset, we compare cross-attention, pairwise self-attention, and bottleneck attention with four pre-trained 3D image encoders. Cross-attention fusion achieves superior performance, with mean ROC-AUC scores of 0.903 (±0.033) for AD vs. cognitively normal (CN) and 0.692 (±0.061) for mild cognitive impairment (MCI) vs. CN. Despite its strong performance, our model remains highly efficient, with only 1.56 million parameters—over 40 times fewer than ResNet-34 (63M) and Swin UNETR (61.98M). These findings demonstrate the potential of cross-attention fusion for improving AD diagnosis while maintaining computational efficiency.

\end{abstract}
\section{Introduction}

Alzheimer's disease (AD) is the most common cause of dementia, leading to irreversible neurodegeneration that ultimately affect patients' daily function and places a significant burden on caregivers and society~\cite{noauthor_2023_2023}. With no cure currently available, early diagnosis before irreversible brain structure change occur is crucial for potential interventions to delay disease progression, manage symptoms and improve patient outcomes. 
Brain atrophy, a hallmark of AD, has been suggested as a potential biomarker for early AD detection, as alterations in brain structure may precede years before noticeable cognitive symptoms~\cite{ridwan_development_2021}. Structural magnetic resonance imaging (sMRI) is the primary imaging modality for assessing brain atrophy and neurodegenerative changes due to its superior tissue contrast and high spatial resolution~\cite{chen_iterative_2021}. Consequently, sMRI has been intensively explored for early AD diagnosis, particularly with the advancements of modern machine learning and computational power.


Deep learning methods based on sMRI, particularly convolutional neural networks (CNN), have shown promise in the diagnosis of AD~\cite{chen_iterative_2021,el-assy_novel_2024,litjens_survey_2017}. However, current approaches face several challenges. First, the majority of current deep learning studies in AD rely on 2D MRI slices, requiring manual selection of informative cross-sections, which may introduce bias and limit the model's ability to capture 3D neurodegenerative patterns. Second, these methods predominantly rely on intensity-based features, which contain rich structural information but often require large sample sizes to effectively capture subtle, localized deformation features -- particularly in 3D whole-brain analysis. 
Third, a few studies incorporate other imaging modalities, such as PET, or clinical and genetic factors like age and APOE status, to improve AD classification~\cite{feng_medai_2023,mustafa_diagnosing_2023,qiang_diagnosis_2023,spasov_parameter-efficient_2019}. While these multi-modal approaches can enhance diagnostic accuracy, PET scans are often unavailable in routine clinical settings due to cost and accessibility constraints, and clinical assessments, including cognitive tests and genetic screening, typically occur later in the diagnostic process. As a result, relying on these features may lead to overly optimistic performance estimates and inflate the true predictive value of MRI-based models for early AD diagnosis.

Alternatively, Jacobian determinant maps (JSM) provide a compact and effective representation of morphological transformations in the brain by quantifying localized expansion or contraction at each voxel relative to a standard template~\cite{ashburner_voxel-based_2000}. JSMs encode structural deformations, allowing for efficient feature extraction and potentially improved sensitivity to neurodegenerative changes, particularly with moderate sample sizes as often seen in AD cohort studies. For instance, hippocampal atrophy in AD manifests not only as changes in signal intensity but also as localized tissue contraction, which JSMs can directly quantify~\cite{van_de_pol_hippocampal_2006}. Previous studies have trained CNN models using whole-brain Jacobian domain features at the subject level to diagnose AD with surprisingly high performance~\cite{qasim_abbas_transformed_2023}. Given these advantages, integrating JSM into deep learning frameworks could potentially enhance sensitivity to structural alterations, particularly in 3D MRI studies with moderate sample size. 


Only a few studies have explored the integration of deformative characteristics with sMRI intensity maps for AD diagnosis. Most existing approaches treat modalities independently, either by concatenating the input level, where the raw sMRI and Jacobian maps are merged at the input stage and fed into a shared feature extractor, or by employing conventional early or late fusion strategies~\cite{qasim_abbas_transformed_2023,mustafa_diagnosing_2023}. 
However, such approaches fail to account for the intrinsic relationship between sMRI and JSM, which originates from the corresponding sMRI scan and provides complementary information about structural changes. 

In this paper, we explore the use of attention mechanisms~\cite{vaswani_attention_2023}, particularly cross-attention~\cite{gheini_cross-attention_2021}, as a framework for modeling intrinsic interactions between sMRI and JSM. We leverage the cross-attention mechanisms by employing JSM to detect deformation regions and direct the extraction of anatomically meaningful sMRI intensity patterns linked to AD.  Although cross-attention has been successfully applied to integrate multimodal imaging data, such as sMRI and PET ~\cite{feng_medai_2023}, for AD diagnosis, to our knowledge, it has not been previously utilized for fusion of JSM and sMRI in AD classification. Given the complementary nature of these modalities -- where sMRI provides intensity-based structural information and JSM encodes localized deformation patterns -- the attention mechanism offers a potential approach to effectively model their interdependencies while preserving their distinct characteristics. 

To evaluate the feasibility of this fusion strategy, we utilize the Alzheimer’s Disease Neuroimaging Initiative (ADNI) database~\cite{jack_jr_alzheimers_2008} and employ a simple CNN encoder integrated with cross-attention. While transformer-based architectures are increasingly explored in medical imaging, CNNs remain well-suited for feature extraction due to their strong spatial inductive biases and relatively low data requirements. Given the moderate sample sizes typically available in AD datasets, a CNN backbone enables us to examine the benefits of cross-attention without introducing excessive computational complexity or requiring large-scale pertaining. This study provides an initial assessment of cross-attention fusion for JSM-MRI in AD classification. By leveraging the complementary strengths of both modalities, we explore whether this integration enhances the model’s ability to distinguish between AD, mild cognitive impairtment (MCI), and cognitive normal (CN) subjects. Our findings offer insights into improving the accuracy and reliability of early AD diagnosis, as well as contributing to the broader development of multimodal deep learning techniques for neuroimaging.

\section{Methods}

\subsection{Data preparation}

The study used T1-weighted sMRI data from the ADNI database~\cite{jack_jr_alzheimers_2008}.
A total of 818 subjects from ADNI-1 were included, including 224 patients with AD, 362 cases of MCI, and 232 controls of CN. Inclusion criteria were: (i) 55 years of age, (ii) diagnosis of CN, AD, and MCI is closet date within 6 months of image acquired date, and (3) availability of high-resolution (1×1×1 mm$^3$) sMRI scans. Each participant was represented by their earliest available image. Demographic and clinical characteristics are summarized in Table~\ref{tab:demographics}.

\begin{table}[H]  
\caption{Demographics and MMSE$^a$ Scores by diagnosis group.}\label{tab:demographics}
\centering
\begin{tabular}{|l|r|r|r|r|r|}
\hline
\textbf{Group} & \textbf{n} & \textbf{Avg. Age} & \textbf{Avg. MMSE Score} & \textbf{Male} & \textbf{Female} \\
\hline
Cognitive Normal (CN) & 232 & 76.44 & 29.03 & 121 & 111 \\ 
\hline
Mild Cognitive Impairment (MCI) & 362 & 75.54 & 26.82 & 234 & 128 \\
\hline
Alzheimer's Disease (AD) & 224 & 75.58 & 22.62 & 121 & 103 \\
\hline
\end{tabular}
\scriptsize{$^a$MMSE denotes Mini-Mental State Examination}
\end{table}
Although these images had already undergone gradient wrap and B1 N3 nonuniformity corrections, residual intensity inhomogeneities were further mitigated using the N4ITK bias field correction algorithm~\cite{tustison_n4itk_2010} as implemented in ANTs (Advanced Normalization Tools)~\cite{avants_symmetric_2008}. Non-brain tissues were removed using SynthStrip~\cite{hoopes2022}, a deep learning-based tool for robust skull stripping across anatomical variability. The skull-stripped, bias-corrected images were first affine-registered to the MNI152 template using ANTs, with a spatial resolution of $182 \times 218 \times 182$ voxels, followed by non-rigid registration with the SyN algorithm~\cite{avants_symmetric_2008}.
The parameters included a step size of 0.1 gradient, multi-resolution optimization (3 levels), and symmetric normalization (SyN) for diffeomorphic transformations~\cite{han_diffeomorphic_2022}. 

During non-linear registration, the deformation field is represented by $V:\mathbb{R}^3 \to \mathbb{R}^3$, which maps the voxel coordinates in the affine-registered space, $X_r=(i_r,j_r,k_r)$, to their corresponding positions in the SyN-registered space, $X_s=(i_s,j_s,k_s)$. The deformation field $V$ can be expressed as: $V(X_r) = v_i^r \hat{I} + v_j^r \hat{J} + v_k^r \hat{K}$, where $\hat{I},\hat{J},\hat{K}$ are the unit vectors along the $x$-, $y$-, and $z$-directions, respectively. The \textit{Jacobian matrix}, denoted as $J_v$, characterizes the local deformation gradient and is computed as the first-order partial derivative of $V$ with respect to the SyN-registered coordinates $X_s$:
\begin{equation}
    J_v = \nabla v =
    \begin{bmatrix}
        \frac{\partial v_i^r}{\partial i_s} & \frac{\partial v_j^r}{\partial j_s} & \frac{\partial v_k^r}{\partial k_s} \\
        \frac{\partial v_i^r}{\partial i_s} & \frac{\partial v_j^r}{\partial j_s} & \frac{\partial v_k^r}{\partial k_s} \\
        \frac{\partial v_i^r}{\partial i_s} & \frac{\partial v_j^r}{\partial j_s} & \frac{\partial v_k^r}{\partial k_s}
    \end{bmatrix}
    \label{eq:jacobian}
\end{equation}
The \textit{Jacobian determinant}, denoted as $|J_v| = \det(\nabla v)$, quantifies the changes in the local volume and is computed as the determinant of the Jacobian matrix.
A value of $|J_v| > 1$ indicates local volume expansion, while $|J_v| <1$ indicates contraction. To symmetrize the distribution and enhance interpretability, we compute the log-Jacobian map as follows:
\begin{equation}
    \log(|J_v|) =
    \begin{cases}
        \log(|J_v|), & \quad \text{if } |J_v| \geq 1, \\
        \log\left(\frac{1}{|J_v|}\right), & \quad \text{if } |J_v| < 1.
    \end{cases}
    \label{eq:log_det_jacobian}
\end{equation}

\subsection{Network Architecture and multimodal fusion}


We implemented dual 3D CNN encoders ( inspired by Qasim \& Abbas~\cite{qasim_abbas_transformed_2023}) to extract features $F_{\text{sMRI}} \in \mathbb{R}^{C \times H \times W \times D}$ and $F_{\text{JSM}} \in \mathbb{R}^{C \times H \times W \times D}$,  flattened to ${N \times C}$ where $N = H \times W \times D$. Modality-specific linear projections assigned features to a $d$-dimensional attention embedding space. In addition to cross-attention fusion, we also compared two other attention-driven fusion strategies: vanilla self-attention, bottleneck attention ~\cite{nagrani_attention_2022}. We used affine-registered sMRI to maintain the spatial relationship of the entire brain~\cite{darzi_review_2024}.

\begin{figure}[H]  
\centering
\noindent\includegraphics[width=0.9\textwidth,keepaspectratio]{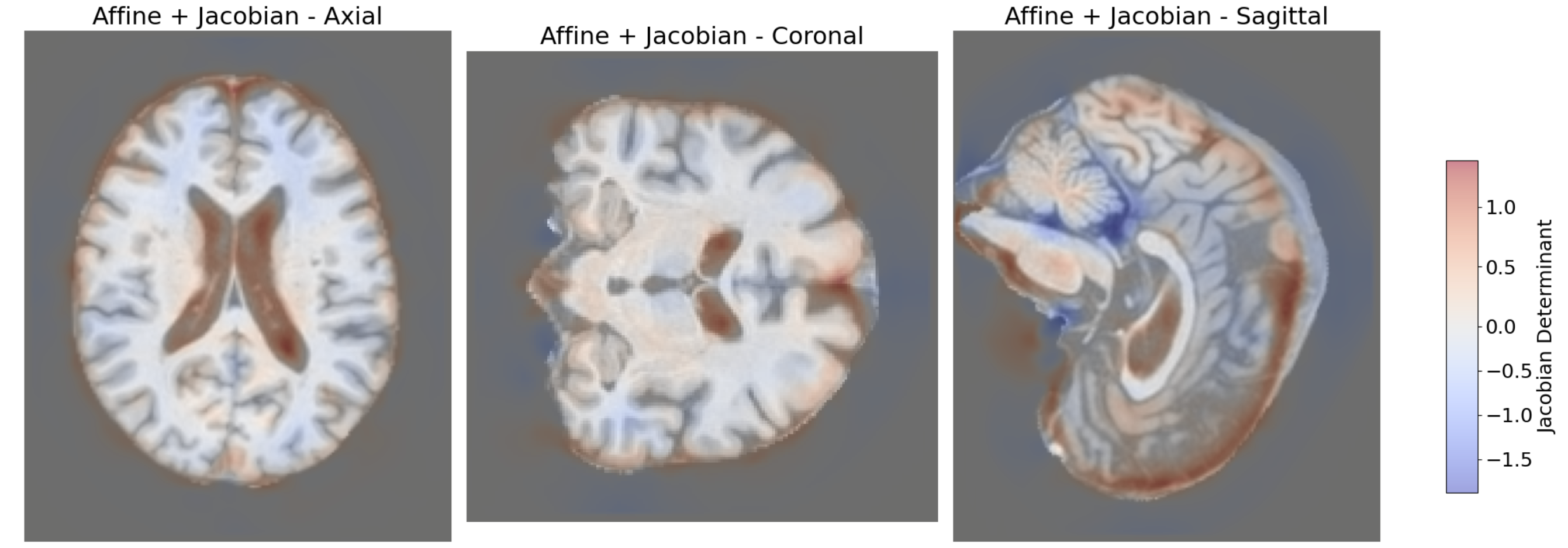}
\caption{Affine Registered sMRI and Jacobian map overlay}
\label{fig:figure2}
\end{figure}

\subsubsection{Self-attention}
We begin with the self-attention fusion method, which contains a multihead self-attention (MSA) module applied  directly to the concatenate sequence of sMRI and JSM features $F_{\text {C }}=\left[F_{\mathrm{sMRI}} \| F_{\mathrm{JSM}}\right] \in \mathbb{R}^{2 N \times d}$. From attention mechanism, each token of $F_{\mathrm{sMRI}}$ and $F_{\mathrm{JSM}}$ can attend within each modality for intramodal attention and each other for cross-modal attention fusion. The query (Q), key (K) and value (V) of are projected from $F_{\text {C}}$

\begin{equation}
    \begin{gathered}
        Z_ {\text {self}} = \text{softmax} \left( \frac{Q_{F_{C}}^\top K_{F_{C}}}{\sqrt{d}} \right)V_{F_{C}}, \quad Z_{\text {self}}\in \mathbb{R}^{2N \times d}\\
    \end{gathered}
    \label{eq:Output_attention}
\end{equation}

After MSA processing, the output $Z_ {\text {self}}=\left[Z_{\text{sMRI}}, Z_{\text{JSM}}\right] \in \mathbb{R}^{2N \times d}$ retains the input dimensions. Global average pooling (GAP) extracts modality-wise means and were added together to form the final feature vector. 


\subsubsection{Bottleneck fusion with Attention}

The bottleneck fusion approach leverages learnable bottleneck tokens, denoted $F_{\mathrm{fsn}} \in \mathbb{R}^{B \times d}$ to mediate cross-modal attention between $F_{\text{sMRI}}$ and $F_{\text{JSM}}$, $B$ is the number of bottleneck tokens shared between modalities~\cite{nagrani_attention_2022}. For each modality, $F_{\mathrm{fsn}}$ were concatenated with modality-specific tokens and applied with MSA to update both modality-specific tokens and bottleneck tokens separately:

\begin{equation}
    \begin{gathered}
        \left[F_i^\prime| F_{\mathrm{fsn}, i}^{\prime}\right]=\operatorname{MSA}\left(\left[F_i\| F_{\mathrm{fsn}}\right] ; \theta_i\right), \quad i \in\{\mathrm{sMRI}, \mathrm{JSM}\}
    \end{gathered}
    \label{eq:Output_attention}
\end{equation}
The bottleneck from both modalities are averaged to produce a shared representations: $F_{\mathrm{fsn}}^{\prime}=\frac{\left(F_{\mathrm{fsn}, \mathrm{sMRI}}^{\prime}+F_{\mathrm{fsn}, \mathrm{JSM}}^{\prime}\right)}{2}$ and the final representations for classification is obtained by average pooled representations of $F_{\text{sMRI}}$ and $F_{\text{JSM}}$ and aggregated bottleneck tokens:
\begin{equation}
    \begin{gathered}
        F_{\text {output }}=\frac{\sum_{i=1}^{N} F_{\mathrm{sMRI}, i}^{\prime}}{N}+\frac{\sum_{i=1}^{N} F_{\mathrm{JSM}, i}^{\prime}}{N}+\frac{\sum_{k=1}^B F_{\mathrm{fsn}, k}^{\prime}}{B} 
    \end{gathered}
    \label{eq:Output_attention}
\end{equation}
Bottleneck fusion restricts cross-modal attention with exclusively few tokens; each model is forced to aggregate the most relevant modal information for fusion and reduces computational complexity due to the small number of bottlenecks~\cite{nagrani_attention_2022}. However, bottleneck fusion compresses sMRI / JSM interactions into restricted latent tokens that potentially oversimplify the structural intensity complexity critical for AD classification.

\subsubsection{Cross-attention}

The proposed framework uses cross-attention to model deformation-structure relationships for Alzheimer’s diagnosis. The deformation patterns of  $F_\text{JSM}$ are encoded as queries $Q_\text{JSM}$ to actively probe the characteristics of  $F_\text{sMRI} $, which are encoded into keys $K_\text{sMRI}$  to filter plausible deformation-structure links and values $V_\text{sMRI}$ that provides tissue-specific intensity information. The scaled dot-product attention computes pairwise correlations between $Q_\text{JSM}$ and $K_\text{sMRI}$, followed by softmax-normalized weighting of $V_\text{sMRI}$, producing cross-attended features that can potentially leverage regions of JSM-derived atrophy to associate sMRI structural changes in the latent space. These features are pooled via GAP and classified by an MLP. 
\begin{equation}
    \begin{gathered}
        Z_{\text{cross}} = \text{softmax}\left(\frac{Q_{\text{JSM}}K_{\text{sMRI}}^T}{\sqrt{d}}\right)V_{\text{sMRI}} \in \mathbb{R}^{N \times d} \\
    \end{gathered}
    \label{eq:Output_attention}
\end{equation}
\subsubsection{Early/late Fusion}
Three baseline fusion approaches were evaluated: (1) Input-Level Fusion (ILF-CNN), where JSM and sMRI are concatenated as input channels and processed by a shared 2-channel CNN; (2) Feature-Level Fusion (FLF-CNN), using modality-specific CNNs to encode and concatenate latent features; and (3) Late Fusion (SC-CNN), employing separate CNNs with logit averaging. Each baseline pairs with an alternate contains intra-modal self-attention (-SA variants) after the CNN outputs. However, these methods treat JSM and sMRI as independent signals, failing to model their intrinsic relationship.


\section{Experiments and Results}
\subsection{Implementation details and evaluation metrics}
To evaluate the effectiveness of architectures, we trained all models under standardized hyper-parameters. Training was performed using Adam Optimizer (learning rate = $1e-4$, default $\beta_{1}$ = 0.9, $\beta_{2}$ = 0.999) with cross-entropy loss and a batch size of 16. The embedded dimensions were set 128 for all attention mechanisms compared. 
We adopted a five-fold cross-validation strategy for all experiments, training each model for 20 epochs per fold. Three binary classification tasks were conducted: (1) CN vs. AD, (2) CN vs. MCI, and (3) MCI vs. AD. All experiments were run on four NVIDIA H100 GPUs. 
We evaluated model performance using the Receiver Operating Characteristic Area Under the Curve (ROC AUC).

\subsection{Performance comparison}

We compared the three proposed multimodal fusion structures (cross-attention, pairwise self-attention and bottleneck attention) with 4 established 3D image encoders ResNet-18/34/50~\cite{chen2019med3d} and Swin UNETR~\cite{tang2022self}, which are pre-trained with large amount of medical images on segmentation tasks. To utilize the pre-trained encoders, we added an MLP classifier and performed full fine-tuning for our classification tasks.


As shown in Table~\ref{tab:simple}, based on the results of 5-folds cross-validation, the cross-attention fusion achieves mean ROC-AUC scores of 0.903 (0.033) and 0.692 (0.061) on the CN vs AD and CN vs. MCI classification tasks respectively, outperforming baseline attention fusion approaches (pairwise self-attention and bottleneck fusion). 
Nevertheless, our approach achieves comparable or superior performance to pre-trained medical image encoders fine-tuned on our tasks, despite having significantly fewer parameters. While 3D ResNet-34/50 each contain 63 and 46 million parameters respectively and Swin UNETR has 61.98 million, our model is remarkably lightweight at just 1.56 million parameters—over 40 times smaller.

Despite having a similar parameter size, the cross-attention fusion strategy outperformed conventional early/late fusion methods and achieved average improvements in ROC-AUC score of 0.067 (CN vs AD), 0.018 (CN vs MCI), and 0.023 (MCI vs AD). Compared to a single-modal identical CNN+MLP trained with sMRI and JSM, the cross-attention fusion strategy improved the mean ROC-AUC by 0.054 (CN vs AD), 0.047 (CN vs MCI) and 0.067 (MCI vs AD), with identical CNN encoder architecture.

\begin{table}[H]
\scriptsize
\centering
\caption{Affine-Registered Model Performance (ROC AUC)}
\label{tab:simple}
\begin{tabular}{@{} l l *{6}{c} c @{}}
\toprule
 & & \multicolumn{2}{c}{CN vs AD} & \multicolumn{2}{c}{CN vs MCI} & \multicolumn{2}{c}{MCI vs AD} & \textbf{Parameters} \\
\cmidrule(lr){3-4} \cmidrule(lr){5-6} \cmidrule(l){7-8} \cmidrule(l){9-9}
\textbf{Category} & \textbf{Model} & \textbf{Mean} & \textbf{Std} & \textbf{Mean} & \textbf{Std} & \textbf{Mean} & \textbf{Std} & (Million) \\
\midrule
\multirow{3}{*}{Attention-Driven Fusion} 
& \textbf{Cross-attention } & \textbf{0.903} & 0.033 &\textbf{ 0.692} & 0.061 & \textbf{0.660} & 0.090 & \textbf{1.56} \\
& BottleNeck       & 0.851 & 0.052 & 0.661 & 0.046 & 0.625 & 0.045 & 1.69 \\
& Self-attention   & 0.826 & 0.061 & 0.654 & 0.060 & 0.622 & 0.044 & 1.56 \\
\hline
\multirow{5}{*}{Early/Late fusion} 
& IFL-CNN  & 0.842  & 0.041 & 0.670  & 0.068 & 0.654 & 0.052 & 0.77 \\
& IFL-CNN-SA  & 0.847   & 0.040 & 0.669   & 0.069 & 0.626 & 0.049 & 0.86 \\
& FLF-CNN& 0.837  & 0.036 & 0.675 & 0.068 & 0.637 & 0.032 & 1.53 \\
& FLF-CNN-SA     & 0.834 & 0.044 & 0.675 & 0.068 & 0.630 & 0.041 & 1.72 \\
& SC-CNN     & 0.833 & 0.048 & 0.689 & 0.055 & 0.636 & 0.048 & 1.53 \\
& SC-CNN-SA      & 0.823  & 0.055  & 0.664  & 0.080 & 0.640 & 0.041 & 1.72 \\
\hline
\multirow{5}{*}{Single Modal} 
& Single sMRI      & 0.849 & 0.064 & 0.645 & 0.047 & 0.625 & 0.107 & 0.77 \\
& Single JSM\cite{qasim_abbas_transformed_2023} & 0.844 & 0.046 & 0.660 & 0.045 & 0.643 & 0.016 & 0.77 \\
& 3D-ResNet18\cite{chen2019med3d}  & 0.876 & 0.050 & 0.690 & 0.048 & 0.646 & 0.068 & 32.99 \\
& 3D-ResNet34\cite{chen2019med3d}  & 0.876 & 0.061 & 0.678 & 0.121 & 0.661 & 0.051 & 63.31 \\
& 3D-ResNet50\cite{chen2019med3d}  & 0.888 & 0.042 & 0.663 & 0.051 & 0.662 & 0.063 & 46.21 \\
& Swin UNETR\cite{tang2022self}   &\textbf{0.895} & 0.016 & \textbf{0.697} & 0.067 &\textbf{ 0.679} & 0.030 & \textbf{61.98} \\
\bottomrule
\end{tabular}

\smallskip
\end{table}

\section{Conclusions and Discussion}

The integration of sMRI and JSM, which was generated as intermediate outputs of SyN registration, offers a framework for improving AD classification using complementary information without requiring additional modalities. While sMRI preserves critical intensity patterns (e.g. gray matter density), JSM quantifies localized structural deformations (e.g., atrophy), together capturing the latent features of both tissue integrity and volumetric changes. 
We proposed a mechanism of cross-attention that allows context-sensitive fusion of sMRI and JSM features in latent space. Unlike conventional self-attention, which indiscriminately models joint feature distributions, cross-attention adaptively models interactions between modalitie's latent features and allows efficient parameter use in making predictions. JSM-derived deformation features act as queries to dynamically weight relevant intensity features from sMRI. This approach not only focuses on extracted features that potentially reflect anatomically meaningful correlations in the latent space, but also reduces redundancy in parameter usage, enhancing computational efficiency.

By combining a cross-attention module with a lightweight CNN encoder, the model achieves diagnostic performance comparable to larger pre-trained architectures, even when trained on limited data. This multimodal structure requires less computational resources, making it suitable for clinical environments where access to advanced infrastructure is restricted. Furthermore, the lightweight of the model allows for localized deployment, reducing the reliance on extensive data sharing while maintaining robustness in diagnosis. The effective fusion of JSM features and sMRI intensity holds promise not only in AD but also in other medical image analysis.

 \bibliographystyle{splncs04}
 \bibliography{reference}
\end{document}